# CryoCiM: Cryogenic Compute-in-Memory based on the Quantum Anomalous Hall Effect

Shamiul Alam[1], Md Mazharul Islam[1], Md Shafayat Hossain[2], Akhilesh Jaiswal[3], and Ahmedullah Aziz[1*]

[1]Dept. of Electrical Eng. and Computer Sci., University of Tennessee, Knoxville, TN, 37996, USA
[2]Dept. of Physics, Princeton University, Princeton, NJ, 08544, USA
[3]Dept. of Electrical and Computer Eng., University of Southern California, Los Angeles, CA, 90089, USA
[*]Corresponding Author. Email: aziz@utk.edu

*Abstract-* **The scaling of the already-matured CMOS technology is steadily approaching its physical limit, motivating the quest for a suitable alternative. Cryogenic operation offers a promising pathway towards continued improvement in computing speed and energy efficiency without aggressive scaling. However, the '*memory wall*' bottleneck of the traditional *von-Neumann* architecture persists even at cryogenic temperature. That is where a compute-in-memory (CiM) architecture, that embeds computing within the memory unit, comes into play. Computations within the memory unit help reduce the expensive data transfer between the memory and the computing units. Therefore, CiM provides extreme energy efficiency that can enable lower cooling cost at cryogenic temperature. In this work, we demonstrate CryoCiM, a cryogenic compute-in-memory framework utilizing a non-volatile memory system based on the quantum anomalous Hall effect (QAHE). Our design can perform memory read/write, and universal binary logic operations (NAND, NOR, and XOR). We design a novel peripheral circuit assembly that can perform the read/write, and single-cycle in-memory logic operations. The utilization of a QAHE-based memory system promises robustness against process variations, through the usage of topologically protected resistive states for data storage. CryoCiM is the first step towards utilizing exclusively cryogenic phenomena to serve the dual purpose of storage and computation with ultra-low power (~ nano-watts) operations.**

Thanks to the continuous improvement in the electronics industry, the matured complementary metal-oxide-semiconductor (CMOS) technology now allows aggressive scaling (evident from the recent demonstration of 2 nm process technology by IBM [1]) and thus, enables the integration of more than a trillion transistors on a single chip (e.g. Cerebras Wafer Scale Engine 2 with 2.6 Trillion transistors [2]). However, the future scalability of CMOS technology will continue to become more challenging due to the exorbitant power dissipation resulting from the aggressive downsizing. This limitation is one of the major concerns in designing high-performance systems at the architecture level. Cryogenic electronics, where the processing and storage units are cooled to very low temperatures (4 K or below), is an extreme yet one of the most promising solutions to circumvent this high-power dissipation issue. Cryogenic operations lead to significant improvement in computing speed and energy efficiency compared to the room temperature operations [illustrated in Fig. 1(a)] [3–5].

Recent studies show that the cryogenic superconductive systems consume around 80 times less energy compared to the room-temperature 7 nm CMOS technology even including the cost required for helium cooling [6]. Nevertheless, the cooling cost remains a challenge for the cryogenic systems, calling for a low-power and low cooling-cost cryogenic system architecture [7]. For low-power operations, the compute-in-memory (CiM) architectures, where computation and memory operations are performed within a memory array, have been explored for cryogenic and room temperatures with both volatile (SRAMs [8–10] and DRAMs [9,11,12]) and non-volatile (RRAMs [13–15], MRAMs [16–19], FeRAMs [20–22], PCMs [23,24], and others [25]) memory platforms. CiM architectures avoid the power-consuming data movement between the processing and memory units and hence, can operate with a very high energy efficiency [26].

Along with the excellent energy efficiency, CiM architectures can lead to throughput improvement. '*Memory wall*', the growing disparity of speed between the processing and memory units, is one of the major challenges for the *von-Neumann* architecture [shown in Fig. 1(b)] in today's big data computing era. With the emergence of data-intensive applications (such as machine learning and artificial intelligence, etc.), the performance mostly depends on the memory units. Moreover, the data transfer between the memory and the processing units causes delay and requires more energy than the computation process itself, posing a bottleneck [27]. For example, *Google's* recent studies show that about 20-42% of the energy is required to drive the high-speed data bus connecting the processing and memory units [28,29]. On the other hand, CiM architectures aim to significantly reduce the data transfer [Fig. 1(c)] and hence, emerge as an attractive solution to the '*memory wall*' bottleneck for data-intensive applications. Also, CiM architectures take advantage of the large internal memory bandwidth and hence solve the issue of limited throughput through massive parallelism.

Beyond traditional computing, quantum computers will also benefit from the CiM architectures. A quantum computing system also comprises several classical hardware components, including a control processor and a memory block [30,31]. The state-of-the-art quantum computing systems necessitate that the control processor and the memory block operate at cryogenic temperature. The speed and energy demands of the existing cryogenic memory systems are not yet comparable to the cryogenic single flux quantum (SFQ) control processors [4,30,32–35]. Thus, an equivalent '*memory wall*' problem arises even in the quantum computing paradigm. Cryogenic CiM can help alleviate this disparity and reduce classically imposed overhead in quantum computing systems [30].

As mentioned earlier, almost all of the volatile and non-volatile memory platforms have been explored for in-memory computing. Such CiM architectures, at room temp, have convincingly demonstrated significant mitigation of memory wall and show better performance with higher throughput and energy efficiency. At cryogenic temperatures, few works including volatile SRAM [10] and non-volatile STT-MRAM [19]-based

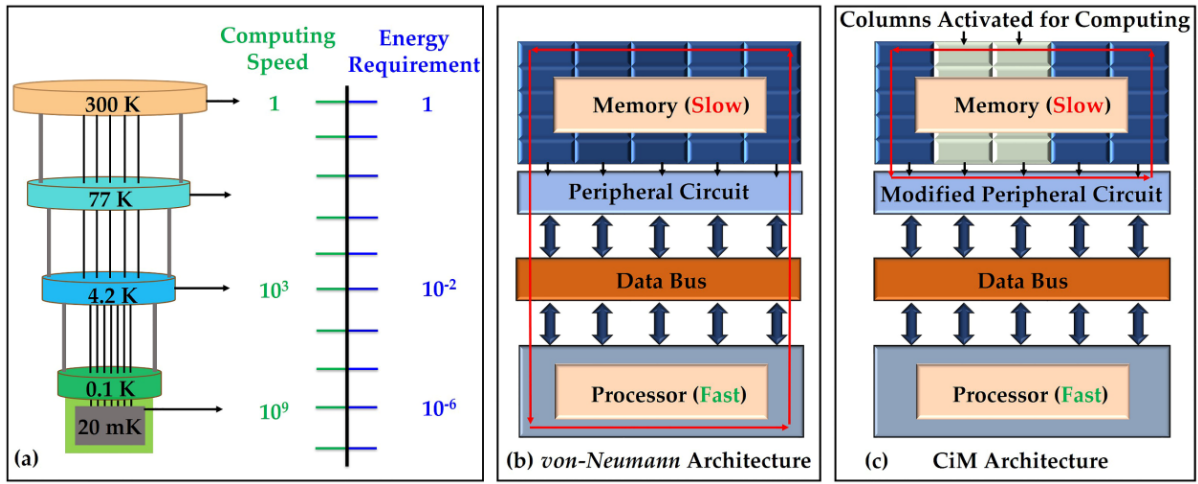

**Fig. 1: (a)** Significant improvement in computing speed and energy requirement with the lowering of operating temperature. Computing speed and energy requirement have been normalized by the value at room temperature (300 K). **(b)** Illustration of the traditional *von-Neumann* architecture which suffers from the data transfer between the memory and the processing units. **(c)** Illustration of the compute-in-memory (CiM) architecture where the modified peripheral circuitry allows the computation within the memory array which solves the memory wall bottleneck.

CiMs have been characterized. It is now well established that the performances (latency and energy efficiency) of the CMOS devices (SRAMs and DRAMs) improve significantly at cryogenic temperatures [3,36] which is supposed to reduce the '*memory wall*' bottleneck. However, the cryogenic temperatures are also supposed to increase the speed of the processing units by about 30-40% [4] which will increase the memory data access rate and worsen the '*memory wall*' issue. Compared to the volatile platforms, the non-volatile platforms show significantly higher energy efficiency at room temperatures [11,37], which makes them very attractive for the cryogenic CiMs. However, in the only cryogenic characterization of non-volatile memory (STT-MRAM)-based CiM architecture [19], it has been reported that the cryogenic operations improve the latency but lower the energy efficiency compared to the room temperature operations. The characterization of FeFET at cryogenic temperatures also reported higher energy requirements compared to the room temperature since the program and erase voltages increase at cryogenic temperature [38].

While researchers, to date, have only characterized the room temperature memory-based CiM architectures at cryogenic temperatures, we report a CiM architecture based on a cryogenic device. Here, we utilize the quantum anomalous Hall effect (QAHE) observed in the twisted bilayer graphene (tBLG) moiré heterostructures [39]. In our previous work [40], we have already reported a cryogenic memory system harnessing the QAHE phenomenon, which provides ultra-low-power performance, high scalability, excellent distinguishability between two logic states, simple peripheral circuitry, and higher sense margin. In this work, we utilize the same memory array with modified peripheral circuitry to demonstrate compute-in-memory. The key contributions made by our work is highlighted below-

- This is the first attempt to design a cryogenic CiM technology using a dissipation-less topological material system that we argue to be a natural fit for this application.
- In contrast to previous works that aimed to characterize conventional CiM technologies at cryogenic temperature [10,19], we design this Cryo-CiM technique from scratch using physical phenomena native to cryogenic environment. Thus, we avert the limitations of the conventional technology platforms.
- Unprecedentedly, our architecture can compute all fundamental logic operations (NAND, NOR, XOR) within a single-cycle. We construct a novel peripheral assembly with dynamically customizable reference to facilitate on-chip sensing and interpretation of different Hall voltage levels which allows the design to achieve this feat.
- Our design continuously computes an analog sum of Hall voltages which dramatically improves performance and reduces power demand.

Having outlined the key aspects and novelty of the design, we discuss the basic building block of our design- QAHE phenomenon and the design of a memory system out of it (reported in Ref. [40]). QAHE, the quantization of Hall resistance in the absence of an external magnetic field, results from the interplay between topological and ferromagnetic properties of electronic band structures. QAHE has been observed in different topological materials [41–47]. Here, we utilize the electrically tunable quantized Hall resistance states observed in tBLG on hexagonal boron nitride (hBN) moiré heterostructure. Figure 2(a) shows the schematic of the heterostructure where bilayer graphene (twisted by an angle of 1.15°) is encapsulated between hBN flakes [48]. Figure 2(b) shows the hysteretic dependence of the Hall resistance on the bias current without any external magnetic field at 4 K. Here, it is evident that the Hall resistance is quantized to two levels ($-h/e^2$ and $h/e^2$) which are used to define the logic states (logic '0' and logic '1', respectively) in the memory cell. Two important bias current levels ($I_{C-}$ and $I_{C+}$) are marked to distinguish the switching points from one memory state to another. These two current levels also separate three regions for three memory operations: (i) $I_{Bias} \leq I_{C-}$: write '0' region, (ii) $I_{Bias} \geq I_{C+}$: write '1' region, and (iii) $I_{C-} < I_{Bias} < I_{C+}$: read region. In the CiM framework (discussed later), we utilize the read region to perform several logic operations. In our simulation, we use the Verilog-A-based phenomenological compact model for the QAHE in tBLG heterostructure which was developed for our previous work [40]. Figure 2(c) shows the behavioral representation of the modeling approach. Figure 2(d) shows the trend for the Hall voltage, $V_{xy}$ (= $I_{Bias} \times R_{xy}$) which elucidates that during the read operation, logic '0' and '1' lead to $V_{xy}$ with opposite signs. As illustrated in Fig. 2(d), for a negative read current, logic '0' and '1' entail positive and negative $V_{xy}$, respectively. This clear distinction in $V_{xy}$ during the read operation makes the sensing and the in-memory logic operations straightforward. In the QAHE-based crosspoint memory system, we utilize the Cu-containing mixed-ionic-electronic-conduction (MIEC) material-based selector [49–51] to successfully write into and read from any specific cell of the array. Figure 2(e) depicts the structure of one cell in the array which consists of one MIEC-based selector and one QAHE-based memory element. For the MIEC selector, we use the look-up table (LUT)-based model developed in Ref. [40]. Figure 2(f) shows the *I-V* characteristics of the MIEC-based selector along with the matching with the experimental data [49–51].

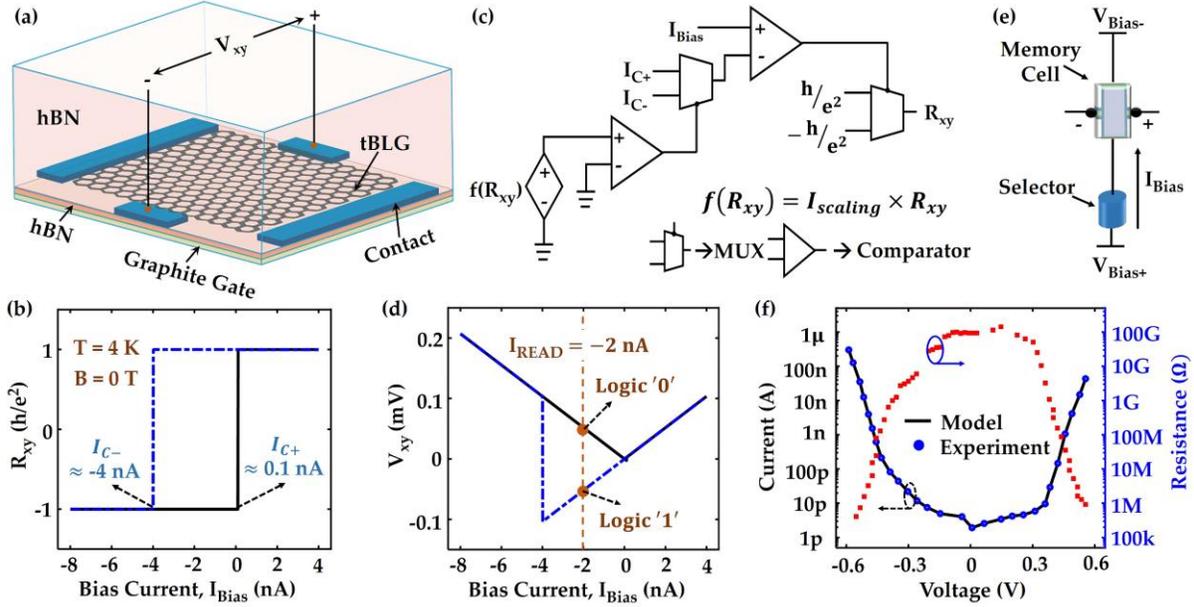

**Fig. 2:** **(a)** Schematic of the tBLG moiré heterostructure where the tBLG is encapsulated between hBN flakes and few-layer graphite flake is used as the gate. **(b)** Hall resistance, $R_{xy}$ as a function of the bias current without any external magnetic field at 4 K. The characteristics shows the electrically tunable hysteretic resistance switching. **(c)** Behavioral representation of the modeling approach for the QAHE observed in the tBLG moiré heterostructure. **(d)** Hall voltage, $V_{xy}$ as a function of the bias current. **(e)** Cell structure of the QAHE-based memory system, containing a QAHE-base memory cell and a mixed-ionic-electronic-conduction (MIEC) material-based selector. **(f)** *I–V* characteristics obtained from a look-up table (LUT) based model of the MIEC-based selector with the experimental matching.

We utilize the crosspoint memory array developed using the MIEC material-based selector and QAHE-based memory element to perform in-memory logic operations. Note, in the previous work, to sense the memory state of a cell in the memory array, a simple voltage comparator, able to differentiate between positive and negative voltage, was sufficient. However, for the in-memory logic operations of this work, a more complicated peripheral circuitry is required. Figure 3(a) shows the complete crosspoint memory array with the modified peripheral circuitry for in-memory computing. In this work, we discuss the read and in-memory logic operations in this memory array. Please refer to our previous work [40] for a detailed discussion on the biasing schemes for accessing a specific cell in the array, write and read operations, and design of peripheral circuitry for the memory operations.

In this memory array, to activate a cell for read and logic operations, a suitable voltage is applied between the bit-line (BL) and word-line (WL) which creates the read current through the corresponding memory cell. As shown in Fig. 2(d), the read current creates two different polarities of $V_{xy}$ for two memory states. We connect the Hall voltage terminals of the memory cells in a row so that the input of the amplifiers of each row gets the sum of the Hall voltages of all the cells in that row [Fig. 3(a)]. The sum of $V_{xy}$ of the activated cells is first amplified from micro-volt level to milli-volt level using a cryogenic voltage amplifier [52–54]. Then this amplified voltage is processed in the sense amplifier which generates the final output for the read and logic operations. The sense amplifier, modified for the CiM purpose, consists of two simple voltage comparators (VC1 and VC2), one inverter, and one AND gate [shown in Fig. 3(b)]. The two voltage comparators are used to compare the amplified voltage with two different reference voltages ($V_{REF1}$ and $V_{REF2}$) set by the reference generator. In our simulation, we utilize a behavioral model for the voltage comparators which behaves as (similar to all the conventional voltage comparators):

$$V_{out} = \begin{cases} 0, & V_{in} < V_{REF} \\ V_{DD}, & V_{in} \geq V_{REF} \end{cases} \quad (1)$$

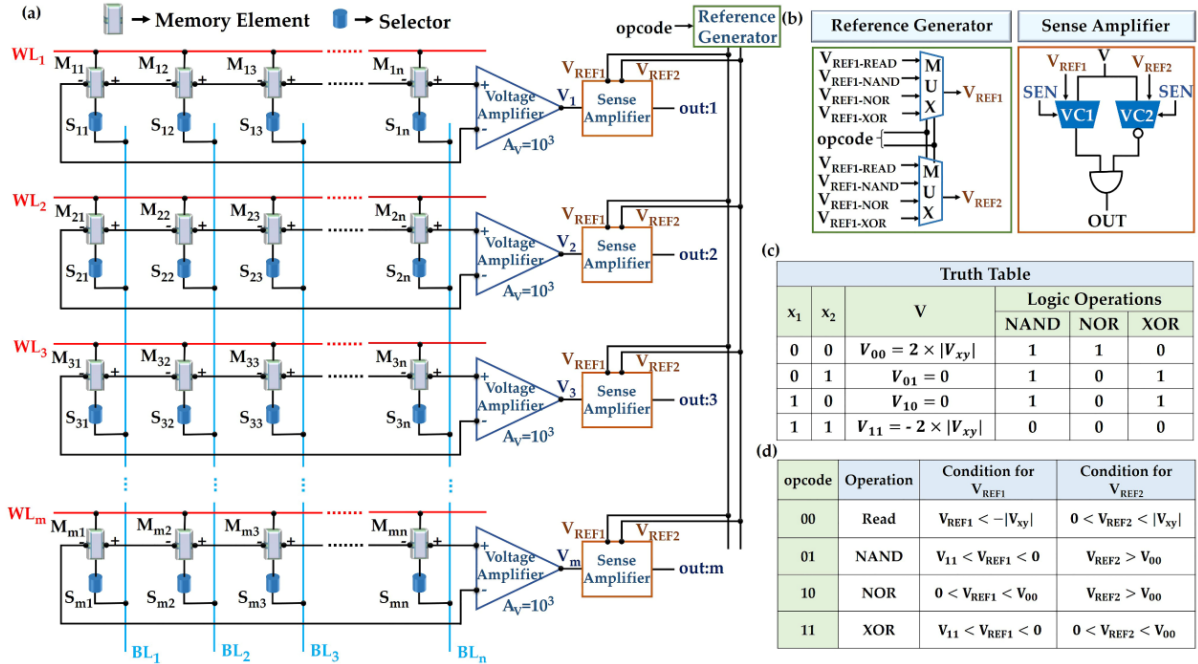

**Fig. 3:** **(a)** QAHE-based cryogenic CiM structure. (Wires of different colors are not connected to each other unless shown using black dots at the intersections between two wires.) **(b)** Schematics of the reference generator and the sense amplifier. **(c)** Truth table for the 2-input logic operations and the corresponding voltage levels for different logic combinations. **(d)** Conditions for choosing the reference voltages for read and in-memory logic operations such as NAND, NOR, and XOR.

However, these voltage comparators can be implemented using the Josephson junction-based voltage comparator [55] and voltage comparators with the CMOS devices characterized at cryogenic temperature [9,56,57]. As shown in Fig. 3(b), the reference generator consists of two 4-to-1 multiplexers (MUX) which generate the two reference voltages for the voltage comparators based on the logic operation to perform (represented by 'opcode'). The table in Fig. 3(d) captures the conditions for choosing the values for the reference voltages to perform different operations (such as memory read, NAND, NOR, and XOR). The noteworthy feature of this sense amplifier design is that it can be configured for different logic operations only by setting suitable reference voltages.

To simulate the read and in-memory logic operations, we develop a framework consisting of a 4 × 4 memory array with the MIEC-based selector, QAHE-based memory element, and the modified peripheral circuitry. Using this framework, we first simulate the memory read operation to demonstrate that the peripheral circuitry, modified for the logic operations, can also be used for the memory read operation. Figure 4(a) shows the biasing scheme to access the first column (keeping all the other columns unaccessed). To demonstrate the read operation for both of the logic states, we utilize the first two cells of the first column (logic '0' in $M_{11}$ and logic '1' in $M_{21}$). Figures 4(c) and 4(d) show the applied read voltage at $BL_1$ and the corresponding Hall voltage generated at the output of the amplifier during read '0' and read '1' operations, respectively. Figures 4(e) and 4(f) show the successful read operation using the peripheral circuitry after activating the sense amplifier and choosing the suitable reference voltages for the read operation. We estimate an average of ~1.75 nW/cell programming power, and ~1 nW/cell read power. Figure 4(b) shows the biasing scheme for the logic operations where we apply the read voltages to the first two columns and assume that the $1^{st}$, $2^{nd}$, $3^{rd}$, and $4^{th}$ rows of the array store '00', '01', '10', and '11' logic conditions, respectively. Figure 4(g) shows the voltage at the output of the amplifiers of the four rows where '00', '01'and '10', and '11' logic combinations entail $\sim 105$ mV, 0 mV, and $\sim -105$ mV at the outputs of the amplifiers, respectively. The clear distinction in the voltage level corresponding to the logic combinations makes the logic operations and the design of peripheral circuitry simple. Figures 4(h), 4(i), and 4(j) present the final outputs for all the logic combinations for NAND, NOR, and XOR operations, respectively. Irrespective of the type of the logic operation, the average power remains ~2 nW per bit per operation (excluding peripheral power).

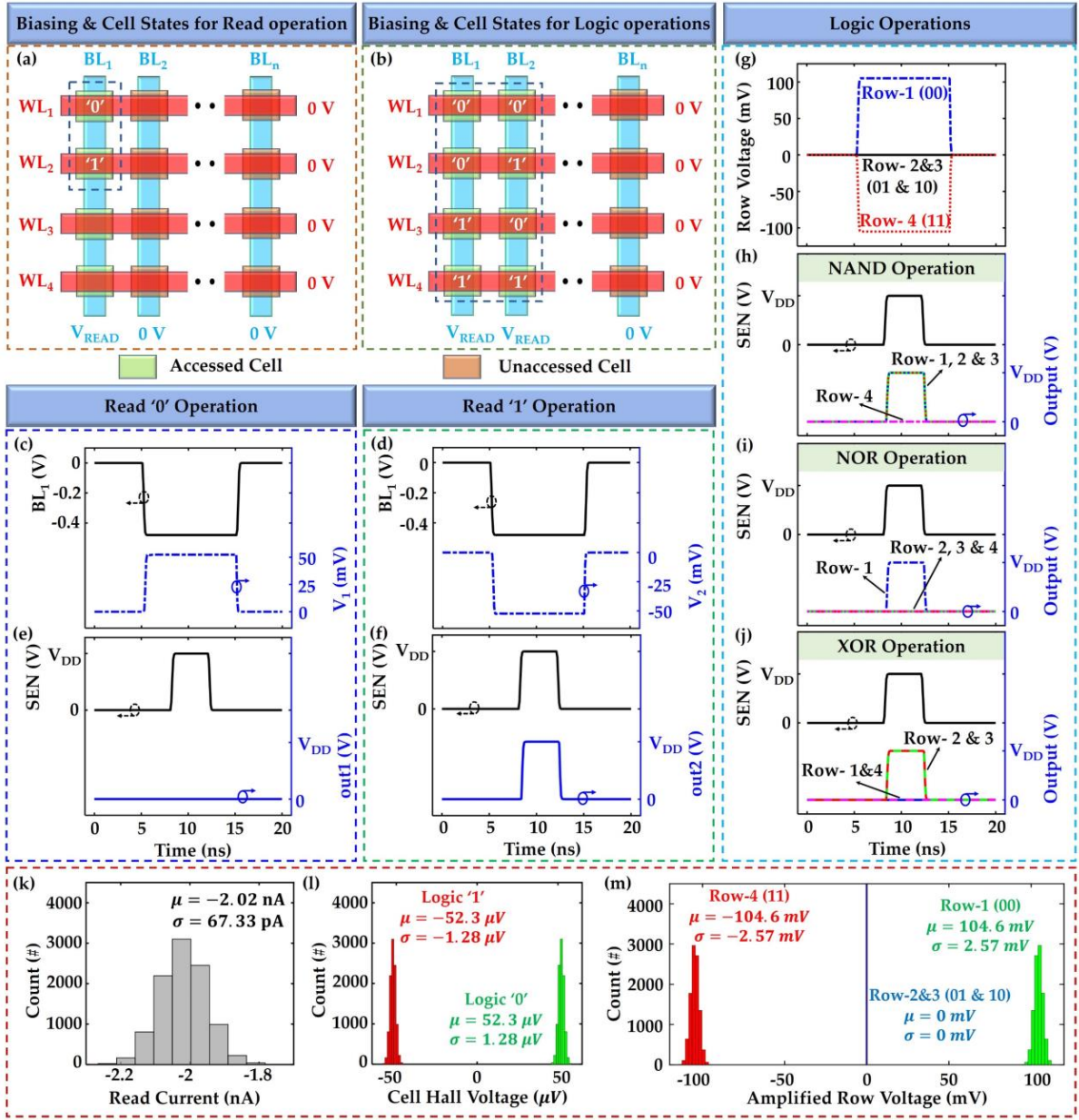

**Fig. 4:** Biasing scheme for the WLs and BLs during **(a)** memory read operation and **(b)** logic operations. Voltage applied to $BL_1$ and the voltage generated at the output of the amplifiers during **(c)** read '0' and **(d)** read '1' operations. Time dynamics of the sense enable signal (SEN) and output of the sense amplifier during **(e)** read '0' and **(f)** read '1' operations. **(g)** Voltage at the outputs of the amplifiers for different logic combinations ('00', '01', '10', and '11') in the rows. Time dynamics of the sense enable signal (SEN) and output of the sense amplifier during **(h)** 2-input NAND, **(i)** 2-input NOR, and **(j)** 2-input XOR operations. **(k)** Distribution of the cell read current with a 10% variation of the nominal value (−2.02 nA) to run a 10000-point Monte-Carlo simulation. Distributions of **(l)** the cell Hall voltage and **(m)** the amplified row voltages (sum of active cell voltages) considering the variation in the read current shown in (k).

We close by discussing the effects of variation on the in-memory logic operations. In terms of variations, QAHE-based memories provide a clear advantage compared to other memories. In all the memories, there are two main sources of variation- (i) the input biases applied to write and read and (ii) the two logic states. Variation from these two sources mostly affects the separation between the memory states which leads to the requirement of complex sensing circuitry and read error in the extreme case. Furthermore, since CiM operations are essentially a modified version of memory read accesses, a reduced sensing margin due to variations inevitably affects the robustness of CiM operations. Advantageously, in QAHE-based memories, the two Hall resistance states are topologically protected and consequently remain constant at $\pm h/e^2$ as long as the temperature remains at 4 K/below [39]. Therefore, there is only one major source of variation in the QAHE-based memories- the variation in the write or read bias. To demonstrate the effects of variation in our QAHE-based in-memory computing, we run a 10000-point Monte-Carlo simulation considering a 10% variation in the read current with its nominal value at -2.2 nA [Fig. 4(k)]. The resulting distribution of the cell Hall voltage is shown in Fig. 4(l). Consequently, we can find the distributions for the amplified row voltages as depicted in Fig. 4(m) which clearly shows that there is no effect of read current variation on the 01/10 logic combinations because the resistance levels remain constant. On the other hand, for '00' and '11' logic combinations, we observe the effect of read current variation in the row voltages. Note that considering the variation, there is still a significant amount of separation between the voltage levels to set the reference voltages. Therefore, the outputs of the in-memory logic operations can be generated using a very simple sense amplifier with two reference voltages.

Cryogenic in-memory computing provides impressive energy efficiency that can solve the *memory wall* bottleneck for both conventional and quantum computing fields and the cooling issues of the cryogenic systems. Here, we have demonstrated a cryogenic CiM based on the QAHE observed in the tBLG moiré heterostructure. We have proposed modified peripheral circuitry with specific biasing schemes along with a reconfigurable sense amplifier for the memory read and in-memory logic operations such as NAND, NOR, and XOR. The sense amplifier can be configured (only suitable reference voltages need to be chosen) to perform different in-memory operations in single-cycle. The QAHE-based CiM provides additional advantages such as- ultra-low-power requirement thanks to the nano-ampere level of switching current, better immunity to variations provided by the topologically protected Hall resistance states, and better scalability enabled by the small cell area and the simpler peripheral circuitry. Through this work, we commence an effort to embed novel CiM techniques in QAHE-based memory systems and possibly beyond.

**Data Availability**

The data that support the plots within this paper and other findings of this study are available from the corresponding author upon reasonable request.

**Author Contributions**

S.A. and A.J. conceived the idea. S.A. performed the array design and the simulations. M.S.H., A.J., and A.A. analyzed the results. A.A. supervised the project. All authors commented on the results and wrote the manuscript.

**Competing Interests**

The authors declare no competing interests.

**References**

[1] B.H. McCarthy and S. Ponedal, IBM News Room 6 (2021).
[2] H. Mujtaba, (n.d.).
[3] N. Yoshikawa, T. Tomida, M. Tokuda, Q. Liu, X. Meng, S.R. Whiteley, and T. Van Duzer, in *IEEE Trans. Appl. Supercond.* (2005).
[4] B. Patra, R.M. Incandela, J.P.G. Van Dijk, H.A.R. Homulle, L. Song, M. Shahmohammadi, R.B. Staszewski, A. Vladimirescu, M. Babaie, F. Sebastiano, and E. Charbon, IEEE J. Solid-State Circuits (2018).
[5] F. Ware, L. Gopalakrishnan, E. Linstadt, S.A. McKee, T. Vogelsang, K.L. Wright, C. Hampel, and G. Bronner, in *ACM Int. Conf. Proceeding Ser.* (2017).
[6] C.L. Ayala, T. Tanaka, R. Saito, M. Nozoe, N. Takeuchi, and N. Yoshikawa, IEEE J. Solid-State Circuits **56**, 1152 (2021).
[7] I. Byun, D. Min, G.H. Lee, S. Na, and J. Kim, Proc. - Int. Symp. Comput. Archit. **2020-May**, 335 (2020).
[8] Z. Jiang, S. Yin, J.S. Seo, and M. Seok, IEEE J. Solid-State Circuits **55**, 1888 (2020).
[9] X. Zhang, V. Mohan, and A. Basu, IEEE Trans. Circuits Syst. II Express Briefs **67**, 816 (2020).
[10] P. Wang, X. Peng, W. Chakraborty, A. Khan, S. Datta, and S. Yu, 1 (2021).
[11] S. Li, C. Xu, Q. Zou, J. Zhao, Y. Lu, and Y. Xie, Proc. - Des. Autom. Conf. **05-09-June**, (2016).
[12] T. Yoo, H. Kim, Q. Chen, T.T.H. Kim, and B. Kim, Proc. Int. Symp. Low Power Electron. Des. **2019-July**, (2019).
[13] NiLeibin, HuangHantao, LiuZichuan, J. V., and YuHao, ACM J. Emerg. Technol. Comput. Syst. **13**, (2017).
[14] S. Yin, J.S. Seo, Y. Kim, X. Han, H. Barnaby, S. Yu, Y. Luo, W. He, X. Sun, and J.J. Kim, IEEE Micro **39**, 54 (2019).
[15] M.K. Bashar, J. Vaidya, A. Mallick, R.S.S. Kanthi, S. Alam, N. Amin, C. Lee, F. Shi, A. Aziz, V. Narayanan, and N. Shukla, (2021).
[16] D. Fan, S. Angizi, and Z. He, Proc. IEEE Comput. Soc. Annu. Symp. VLSI, ISVLSI **2017-July**, 683 (2017).
[17] W. Kang, H. Wang, Z. Wang, Y. Zhang, and W. Zhao, IEEE Trans. Magn. **53**, (2017).
[18] Z. He, S. Angizi, and D. Fan, Proc. - 35th IEEE Int. Conf. Comput. Des. ICCD 2017 439 (2017).
[19] S. Resch, H. Cilasun, and U.R. Karpuzcu, IEEE Comput. Archit. Lett. **20**, 74 (2021).
[20] D. Reis, M. Niemier, and X. Sharon Hu, Proc. Int. Symp. Low Power Electron. Des. **18**, (2018).
[21] K. Ni, B. Grisafe, W. Chakraborty, A.K. Saha, S. Dutta, M. Jerry, J.A. Smith, S. Gupta, and S. Datta, Tech. Dig. - Int. Electron Devices Meet. IEDM **2018-Decem**, 16.1.1 (2019).
[22] W. Qiao, Y. Zhao, J. Yang, C. Liu, P. Jiang, Q. Ding, T. Gong, Q. Luo, H. Lv, and M. Liu, 2020 IEEE 15th Int. Conf. Solid-State Integr. Circuit Technol. ICSICT 2020 - Proc. (2020).
[23] P. Hosseini, A. Sebastian, N. Papandreou, C.D. Wright, and H. Bhaskaran, IEEE Electron Device Lett. **36**, 975 (2015).
[24] I. Chakraborty, G. Saha, and K. Roy, Phys. Rev. Appl. **11**, 014063 (2019).
[25] L. Yin, R. Cheng, Y. Wen, C. Liu, and J. He, Adv. Mater. **33**, 2007081 (2021).
[26] O. Mutlu, 2020 Int. Symp. VLSI Des. Autom. Test, VLSI-DAT 2020 (2020).
[27] V. Sze, Y.H. Chen, J. Einer, A. Suleiman, and Z. Zhang, Proc. Cust. Integr. Circuits Conf. **2017-April**, (2017).
[28] S. Kanev, J.P. Darago, K. Hazelwood, P. Ranganathan, T. Moseley, G.-Y. Wei, and D. Brooks, in *Proc. 42nd Annu. Int. Symp. Comput. Archit.* (ACM, New York, NY, USA, 2015), pp. 158–169.
[29] A. Boroumand, S. Ghose, Y. Kim, R. Ausavarungnirun, E. Shiu, R. Thakur, D. Kim, A. Kuusela, A. Knies, P. Ranganathan, and O. Mutlu, Proc. Twenty-Third Int. Conf. Archit. Support Program. Lang. Oper. Syst. **18**, (2018).
[30] S.S. Tannu, D.M. Carmean, and M.K. Qureshi, in *ACM Int. Conf. Proceeding Ser.* (2017).
[31] S.K. Tolpygo, Low Temp. Phys. (2016).
[32] C. Ahn, S. Kim, T. Gokmen, O. Dial, M. Ritter, and H.S.P. Wong, in *Proc. Tech. Progr. - 2014 Int. Symp. VLSI Technol. Syst. Appl. VLSI-TSA 2014* (2014).
[33] L. Lang, Y. Jiang, F. Lu, C. Wang, Y. Chen, A.D. Kent, and L. Ye, Appl. Phys. Lett. **116**, 022409 (2020).
[34] S. Alam, M.S. Hossain, and A. Aziz, Appl. Phys. Lett. **119**, 082602 (2021).
[35] S. Alam, M.S. Hossain, S.R. Srinivasa, and A. Aziz, arXiv:2111.09436 (2021).
[36] G.H. Lee, D. Min, I. Byun, and J. Kim, Proc. - Int. Symp. Comput. Archit. 774 (2019).
[37] S. Resch, S.K. Khatamifard, Z.I. Chowdhury, M. Zabihi, Z. Zhao, H. Cilasun, J.P. Wang, S.S. Sapatnekar, and U.R. Karpuzcu, Proc. Annu. Int. Symp. Microarchitecture, MICRO **2020-Octob**, 400 (2020).
[38] Z. Wang, H. Ying, W. Chern, S. Yu, M. Mourigal, J.D. Cressler, and A.I. Khan, Appl. Phys. Lett. (2020).
[39] M. Serlin, C.L. Tschirhart, H. Polshyn, Y. Zhang, J. Zhu, K. Watanabe, T. Taniguchi, L. Balents, and A.F. Young, Science (80-. ). **367**, 900 (2020).
[40] S. Alam, M.S. Hossain, and A. Aziz, Sci. Rep. **11**, 1 (2021).
[41] C.Z. Chang, J. Zhang, X. Feng, J. Shen, Z. Zhang, M. Guo, K. Li, Y. Ou, P. Wei, L.L. Wang, Z.Q. Ji, Y. Feng, S. Ji, X. Chen, J. Jia, X. Dai, Z. Fang, S.C. Zhang, K. He, Y. Wang, L. Lu, X.C. Ma, and Q.K. Xue, Science (80-. ). (2013).
[42] C.Z. Chang, W. Zhao, D.Y. Kim, H. Zhang, B.A. Assaf, D. Heiman, S.C. Zhang, C. Liu, M.H.W. Chan, and J.S. Moodera, Nat. Mater. (2015).
[43] K. Yasuda, M. Mogi, R. Yoshimi, A. Tsukazaki, K.S. Takahashi, M. Kawasaki, F. Kagawa, and Y. Tokura, Science (80-. ). (2017).
[44] I. Lee, C.K. Kim, J. Lee, S.J.L. Billinge, R. Zhong, J.A. Schneeloch, T. Liu, T. Valla, J.M. Tranquada, G. Gu, and J.C.S. Davis, Proc. Natl. Acad. Sci. U. S. A. (2015).
[45] Y. Deng, Y. Yu, M.Z. Shi, Z. Guo, Z. Xu, J. Wang, X.H. Chen, and Y. Zhang, Science (80-. ). **367**, 895 (2020).

[46] J.X. Yin, W. Ma, T.A. Cochran, X. Xu, S.S. Zhang, H.J. Tien, N. Shumiya, G. Cheng, K. Jiang, B. Lian, Z. Song, G. Chang, I. Belopolski, D. Multer, M. Litskevich, Z.J. Cheng, X.P. Yang, B. Swidler, H. Zhou, H. Lin, T. Neupert, Z. Wang, N. Yao, T.R. Chang, S. Jia, and M. Zahid Hasan, Nature (2020).

[47] H. Deng, Z. Chen, A. Wolos, M. Konczykowski, K. Sobczak, J. Sitnicka, I. V. Fedorchenko, J. Borysiuk, T. Heider, L. Plucinski, K. Park, A.B. Georgescu, J. Cano, and L. Krusin-Elbaum, Nat. Phys. 2020 1 (2020).

[48] B. Lian, X.Q. Sun, A. Vaezi, X.L. Qib, and S.C. Zhang, Proc. Natl. Acad. Sci. U. S. A. (2018).

[49] K. Gopalakrishnan, R.S. Shenoy, C.T. Rettner, K. Virwani, D.S. Bethune, R.M. Shelby, G.W. Burr, A. Kellock, R.S. King, K. Nguyen, A.N. Bowers, M. Jurich, B. Jackson, A.M. Friz, T. Topuria, P.M. Rice, and B.N. Kurdi, in *Dig. Tech. Pap. - Symp. VLSI Technol.* (2010).

[50] G.W. Burr, K. Virwani, R.S. Shenoy, A. Padilla, M. BrightSky, E.A. Joseph, M. Lofaro, A.J. Kellock, R.S. King, K. Nguyen, A.N. Bowers, M. Jurich, C.T. Rettner, B. Jackson, D.S. Bethune, R.M. Shelby, T. Topuria, N. Arellano, P.M. Rice, B.N. Kurdi, and K. Gopalakrishnan, in *Dig. Tech. Pap. - Symp. VLSI Technol.* (2012).

[51] R.S. Shenoy, K. Gopalakrishnan, B. Jackson, K. Virwani, G.W. Burr, C.T. Rettner, A. Padilla, D.S. Bethune, R.M. Shelby, A.J. Kellock, M. Breitwisch, E.A. Joseph, R. Dasaka, R.S. King, K. Nguyen, A.N. Bowers, M. Jurich, A.M. Friz, T. Topuria, P.M. Rice, and B.N. Kurdi, in *Dig. Tech. Pap. - Symp. VLSI Technol.* (2011).

[52] B.I. Ivanov, M. Trgala, M. Grajcar, E. Ilichev, and H.G. Meyer, Rev. Sci. Instrum. (2011).

[53] T. Arakawa, Y. Nishihara, M. Maeda, S. Norimoto, and K. Kobayashi, Appl. Phys. Lett. (2013).

[54] S. Weinreb, J.C. Bardin, and H. Mani, IEEE Trans. Microw. Theory Tech. (2007).

[55] R.F. Dziuba, B.F. Field, and T.F. Finnegan, IEEE Trans. Instrum. Meas. **23**, 264 (1974).

[56] J. He, T. Teng Lu, G. Guo, J. Xu, and C. Luo, Proc. - 2020 2nd Int. Conf. Inf. Technol. Comput. Appl. ITCA 2020 86 (2020).

[57] E. Charbon, F. Sebastiano, A. Vladimirescu, H. Homulle, S. Visser, L. Song, and R.M. Incandela, Tech. Dig. - Int. Electron Devices Meet. IEDM 13.5.1 (2017).

# Supplementary Material for

# CryoCiM: Cryogenic Compute-in-Memory based on the Quantum Anomalous Hall Effect

Shamiul Alam[1], Md Mazharul Islam[1], Md Shafayat Hossain[2], Akhilesh Jaiswal[3], and Ahmedullah Aziz[1*]

[1]Dept. of Electrical Eng. and Computer Sci., University of Tennessee, Knoxville, TN, 37996, USA
[2]Dept. of Physics, Princeton University, Princeton, NJ, 08544, USA
[3]Dept. of Electrical and Computer Eng., University of Southern California, Los Angeles, CA, 90089, USA
[*]Corresponding Author. Email: aziz@utk.edu

1. **Access Scheme for In-memory Logic Operations:**

In the main text, we briefly discussed the biasing scheme to access two cells per row in the QAHE memory array to perform the compute-in-memory operations. Here we present two example scenarios to demonstrate single-row activation and multi-row activation processes. The CryoCiM array allows independent row-wise access to perform logic operations using the stored data in two cells within the same row. Figure S1 shows the case (I) where all the rows are activated simultaneously and the first two cells (without any loss of generality) in each row are used for logic operations. Figure S2 illustrates the case (II) where only the first row is activated, and the first two cells of that row are used for logic operations. For case II, the cells in the inactive rows will have a non-zero voltage bias ($V_{READ}/2$) across them. It will give rise to leakage current (hence leakage power) across the array but will not lead to any functional failure/read disturb. The selector device (connected in series with the QAHE devices in every cell) helps to suppress the leakage current and prevents any accidental data flip.

Note, in both figures, any overlap between lines with the same color represents a direct connection. But any crossover/overlap between lines of different color does not represent a connection, unless a black dot is placed on the point of intersection. This drawing convention helps identifying the directions of current flow through the memory cells (marked with dashed magenta lines).

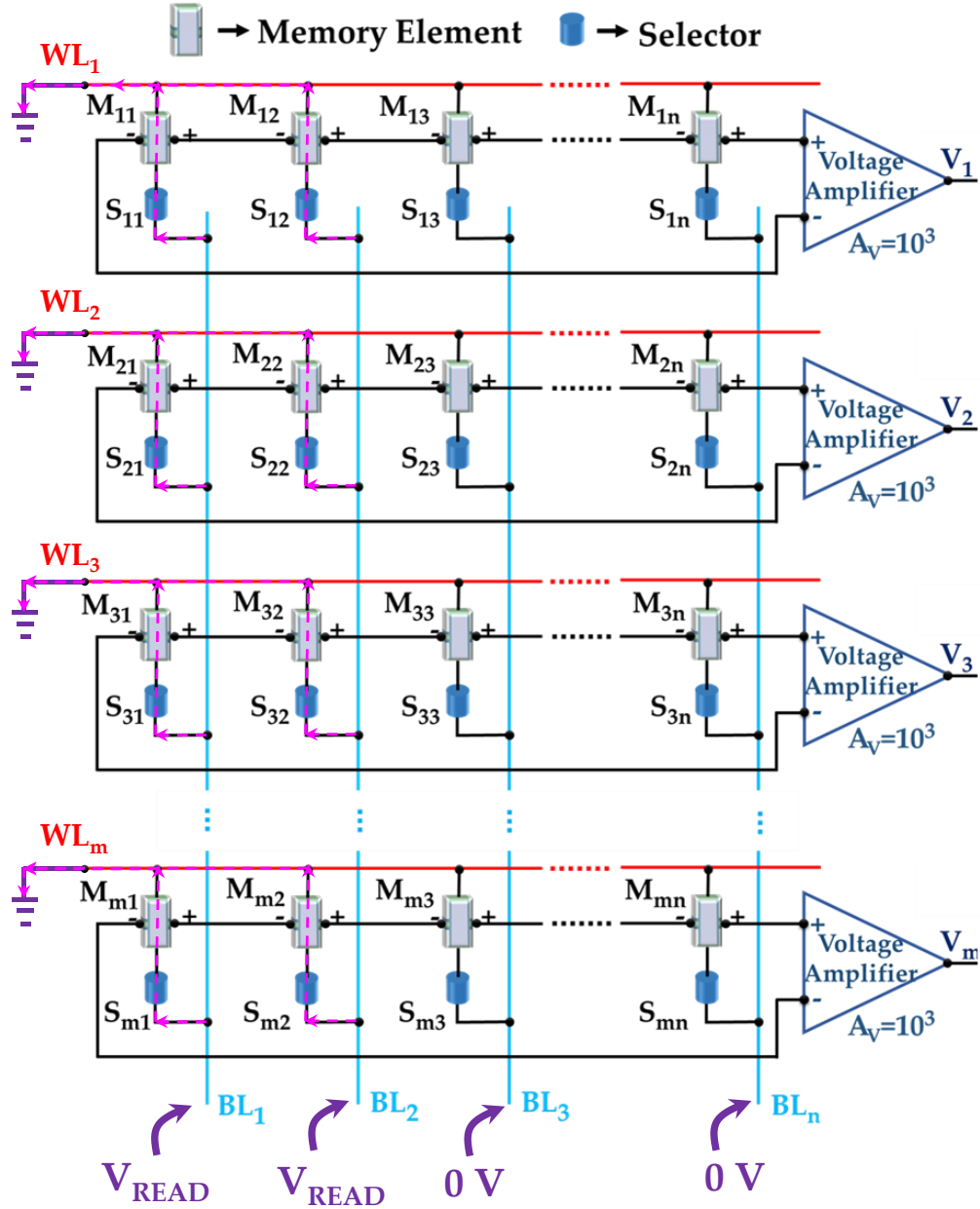

**Fig. S1:** Case I: Biasing scheme and the direction of the current flow (dashed magenta line) through the cells while performing in-memory logic operations. Here, we apply bias voltages to the bit lines to access the first two cells in each row. CryoCiM allows simultaneous multi-row logic operations.

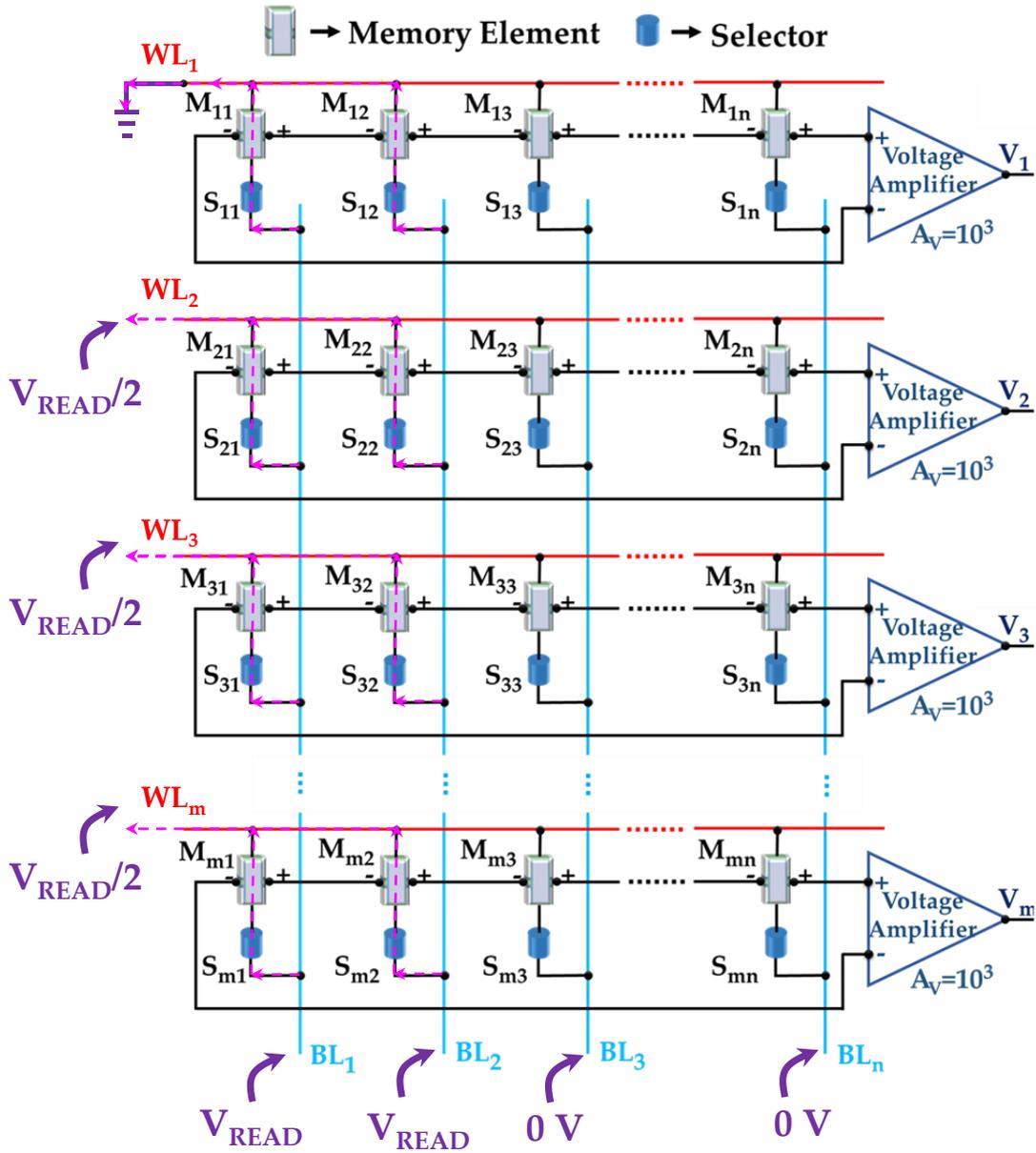

**Fig. S2:** Case II: Biasing scheme and the direction of the current flow (dashed magenta line) through the cells while performing in-memory logic operations. Here, we apply bias voltages to the bit lines to access two cells only in the first row.